\documentclass[12pt, a4paper]{article}
\usepackage{cite}
\usepackage{amsmath,amssymb}
\input{colordvi.tex}
\usepackage[dvipdfmx]{color}
\usepackage[dvipdfmx]{graphicx}
\usepackage{comment}
\begin{document}

%%%%%%%%%%%%%%%%%%%%%%%%%%%%%%%%%%%%%%%%%%%%%%%%%%
\begin{titlepage}

\begin{center}

\hfill UT-18-09\\

\vskip .75in

{\Large \bf Axion-Photon Conversion and Effects on \\ \vspace{2mm} 21cm Observation}

\vskip .75in

{\large Takeo Moroi$^{(a,b)}$, Kazunori Nakayama$^{(a,b)}$ and Yong Tang$^{(a)}$ }

\vskip 0.25in

$^{(a)}${\em Department of Physics, Faculty of Science,\\
The University of Tokyo,  Bunkyo-ku, Tokyo 113-0033, Japan}\\[.3em]
$^{(b)}${\em Kavli IPMU (WPI), UTIAS,\\
The University of Tokyo,  Kashiwa, Chiba 277-8583, Japan}

\end{center}
\vskip .5in

\begin{abstract}

  Recently the EDGES experiment reported an enhanced 21cm absorption signal in the radio wave
  observation, which may be interpreted as either anomalous cooling of
  baryons or heating of cosmic microwave background photons.  In this
  paper, we pursue the latter possibility.  We point out that dark
  radiation consisting of axion-like particles can resonantly convert
  into photons under the intergalactic magnetic field, which can
  effectively heat up the radiation in the frequency range relevant
  for the EDGES experiment. This may explain the EDGES anomaly.

\end{abstract}

\end{titlepage}

%\tableofcontents

\renewcommand{\thepage}{\arabic{page}}
\setcounter{page}{1}
\renewcommand{\thefootnote}{\#\arabic{footnote}}
\setcounter{footnote}{0}
%%%%%%%%%%%%%%%%%%%%%%%%%%%%%%%%%%%%%%%%%%%%%%%%%%

\newpage

%%%%%%%%%%%%%%%%%%%%%%%%%%%%%%%%%%%%%%%%%%%%%%%%%%
\section{Introduction}
\label{sec:Intro}
%%%%%%%%%%%%%%%%%%%%%%%%%%%%%%%%%%%%%%%%%%%%%%%%%%

Recently the EDGES experiment announced a measurement of the cosmic radio background
flux at the frequency range $50$\,MHz--$100$\,MHz to search for the
21cm absorption signal at the epoch of redshift $z\sim 20$,
and found anomalously large absorption than that expected in the standard
reionization models~\cite{Bowman:2018yin}. 
They measured the 21cm brightness temperature relative to that of the cosmic microwave background (CMB), which is given by~\cite{Zaldarriaga:2003du,Furlanetto:2006jb}
\begin{align}
	T_{21}(z) \simeq 23\,{\rm mK}\times x_{\rm HI}(z)
	\left(\frac{\Omega_{\rm b} h^2}{0.02}\right) \left(\frac{1+z}{10}\frac{0.15}{\Omega_{\rm m}h^2} \right)^{1/2}
	\left(1-\frac{T_\gamma(z)}{T_{\rm S}(z)}\right) ,
\end{align}
where $x_{\rm HI}$ is the neutral fraction of the hydrogen atom,
$T_{\gamma}$ is the brightness temperature of the CMB, $T_{\rm S}$ is
the spin temperature, $\Omega_{\rm b}$ and $\Omega_{\rm m}$ are the
density parameters of the baryon and matter and $h$ is the present
Hubble parameter in units of $100\,{\rm km\,s^{-1}\,Mpc^{-1}}$.  It
was realized~\cite{Bowman:2018yin} that the result may indicate that
either the baryon gas temperature is lower than the standard scenario
by a factor $\sim 2$ to reduce the spin temperature by a factor
$\sim 2$,\footnote{ Around the epoch of $z\sim 17$, Lyman $\alpha$ photons
  are produced by stars and they couple the gas temperature to the
  spin temperature~\ \cite{Wouthuysen:1952,Field:1958}.  } or the CMB
brightness temperature at that frequency range is higher by a factor
$\sim 2$.  The former interpretation was achieved by introducing
relatively large scattering of baryon with dark matter so that the
kinetic energy of baryon is transferred to dark
matter~\cite{Barkana:2018lgd}. However, this explanation was soon
challenged and actually highly constrained by cosmological
observations and laboratory experiments~\cite{Munoz:2018pzp,Fialkov:2018xre,
  Berlin:2018sjs, Barkana:2018qrx}. Since then various new constraints
on dark matter from this measurement has also been studied~\cite{Fraser:2018acy, DAmico:2018sxd,
  Clark:2018ghm, Cheung:2018vww, Slatyer:2018aqg, Liu:2018uzy,
  Munoz:2018jwq, Jia:2018csj}. As for the latter possibility, one may introduce
light particles decaying into photons with CMB Rayleigh-Jeans tail
frequency range, but it is found that the required decay rate to
explain the anomaly is so large that it requires some additional
mechanism to enhance the energy injection rate such as
minicluster~\cite{Fraser:2018acy}.

A simple alternative scenario along the latter possibility was considered in Ref.~\cite{Pospelov:2018kdh},
where it is proposed that the hidden photon background, existing as dark radiation, are resonantly converted into photons through the kinetic mixing at the redshift $20 < z < 1700$. Since the number of hidden photons can be many orders of magnitude larger than the CMB photon number density at the Rayleigh-Jeans tail frequency range, even a small conversion rate can lead to a factor 2 enhancement of the CMB brightness temperature.

In this paper, motivated by the EDGES result, we consider axion-like
particle (ALP) dark radiation, which are resonantly converted into 
photons. The existence of ALPs may be ubiquitous in string-theoretic
framework and their masses and decay constants can take wide range of
values~\cite{Jaeckel:2010ni}.  Moreover, there is evidence of the
existence of the primordial magnetic field $B_0 \gtrsim 10^{-17}$\,G
on Mpc scales~\cite{Durrer:2013pga}.  On these grounds, we examine the
possibility that the ALP conversion into photon under the primordial
magnetic fields explains the EDGES anomaly.\footnote{ See also
  Refs.~\cite{Lambiase:2018lhs,Lawson:2018qkc} for some other
  relations between the axion and 21cm signal, and
  Ref.~\cite{Falkowski:2018qdj} for modification of recombination
  history. } Cosmological effects of (resonant) axion conversion into
the photon have been considered in
Refs.~\cite{Yanagida:1987nf,Higaki:2013qka,Evoli:2016zhj}, but the
case of axion dark radiation with energy much lower than the CMB
photon was not considered so far.  We will show that actually the
axion dark radiation with such low energy can modify the CMB spectrum
of its Rayleigh-Jeans tail and hence it can affect the 21cm
observation.

%%%%%%%%%%%%%%%%%%%%%%%%%%%%%%%%%%%%%%%%%%%%%%%%%%
\section{Resonant conversion of axion into photon}
\label{sec:res}
%%%%%%%%%%%%%%%%%%%%%%%%%%%%%%%%%%%%%%%%%%%%%%%%%%

The ALP, represented by $a$, is assumed to have a coupling to photon
as
\begin{align}
	\mathcal L = -\frac{1}{4} g_a a F_{\mu\nu}\widetilde F^{\mu\nu},
\end{align}
where $F_{\mu\nu}$ is the electromagnetic field strength tensor,
$\widetilde F^{\mu\nu}$ is its dual, and $g_a$ represents the strength
of the ALP-photon coupling.  Under the background magnetic field $\vec
B$, it gives an effective mixing between the ALP and
photon~\cite{Raffelt:1987im}.  Let us consider ALP/photon with energy
$E$ which passes through the region with non-vanishing magnetic field.
The effective mass matrix of the ALP and photon looks like
\begin{align}
	\mathcal M^2 = \begin{pmatrix}
		m_a^2 & E g_a B_{\perp} \\
		Eg_a B_{\perp} & \omega_p^2
	\end{pmatrix},
\end{align}
where $B_{\perp}$ denotes the strength of the magnetic field
perpendicular to the ALP/photon momentum direction and we have
included the effective photon mass, the so-called plasma frequency,
\begin{align}
	\omega_p(z) = \sqrt{\frac{4\pi \alpha n_e(z)}{m_e}} \simeq 1.6\times 10^{-14}\,{\rm eV} \,(1+z)^{3/2} X_e^{1/2}. 
\end{align}
Here $\alpha$ is the electromagnetic fine structure constant, $m_e$ is the electron mass, $z$ is the redshift, and $X_e$ denotes the ionization fraction of the hydrogen atom. The mixing angle between photon and ALP is found to be
\begin{align}
	\sin^2(2\theta) = \frac{(2E g_a B_{\perp})^2}{(2Eg_aB_{\perp})^2 + (\omega_p^2- m_a^2)^2}.
\end{align}
Since $\omega_p(z)$ changes with Hubble expansion, there is an epoch,
denoted by $z=z_{\rm res}$, at which $\omega_p^2$ becomes equal to
$m_a^2$ for a certain range of ALP mass.  For the resonant conversion
to happen after the matter-radiation equality, for example, the ALP
mass should be in the following range: $m_a \sim 10^{-14}\,{\rm eV} -
10^{-9}\,{\rm eV}$.  Then the mixing angle becomes maximum and the
resonant conversion of the ALP into photon (or the opposite process)
happens~\cite{Yanagida:1987nf,Mirizzi:2009nq}.  The conversion
probability of the ALP into photon after the resonance
is~\cite{Mirizzi:2009nq}
\begin{align}
	P_{a\to \gamma} = \frac{1}{2} \left[ 1 - \exp\left(-\frac{2\pi r g_a^2 B_{\perp}^2 E}{m_a^2}\right) \right]
	\simeq \frac{\pi r g_a^2 B_{\perp}^2 E}{m_a^2},
	\label{P_res}
\end{align}
where, denoting the expansion rate of the Hubble parameter as $H$, 
\begin{align}
	r^{-1} \equiv \frac{d \ln \omega_p^2}{dt} = 3H + \frac{d\ln X_e}{dt},
\end{align}
with all quantities being evaluated at $z=z_{\rm res}$.  Here, we have
assumed the probability is much smaller than unity in the last
similarity of (\ref{P_res}).  As we will see below, such a condition
holds for the case of our interest.  Note that in a cosmological
setup, the relative direction of $\vec B$ is random against the
ALP/photon momentum direction, and hence $B_{\perp}^2$ in
(\ref{P_res}) should be regarded as $\left< B_{\perp}^2\right> =
B^2/3$.  We assume $B(z) = B_0 (1+z)^2$~\cite{Durrer:2013pga}.

%%%%%%%%
\begin{figure}
	\begin{center}
		\includegraphics[width=0.6\textwidth,height=0.5\textwidth]{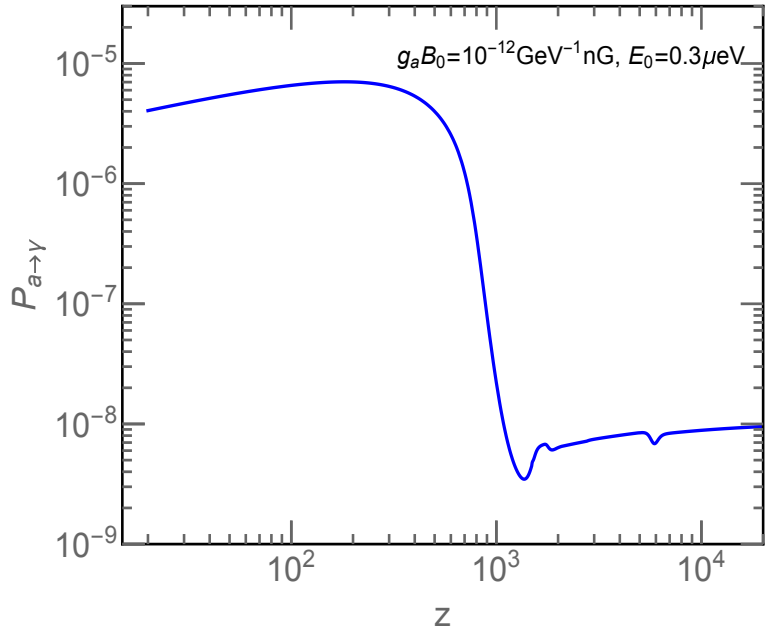}
		\caption{The conversion probability $P_{a\rightarrow\gamma}$ as a function of the redshift at he resonance $z_{\rm res}$. 
		We have used $g_a B_0= 10^{-12}\,$GeV$^{-1}$nG and $E_0=0.3\,\mu$eV for illustration. \label{fig:prop}}
	\end{center}
\end{figure}
%%%%%%%%

In Fig.~\ref{fig:prop} we show the conversion probability
$P_{a\rightarrow\gamma}$ as a function of redshift at the resonance,
$z_{\rm res}$. We have taken $g_a B_0= 10^{-12}\,$GeV$^{-1}$nG and
$E_0=0.3\,\mu$eV for illustration (with $E_0$ being the present energy
of ALP/photon); the EDGES frequency range $50$\,MHz--$100$\,MHz
corresponds to $E_0 \simeq (0.2-0.4)\,\mu$eV.\footnote{ The relation
  between energy $E$ and frequency $f$ is given as $E = 2\pi f \simeq
  4.14\,\mu{\rm eV} (f/1\,{\rm GHz})$.}  We have used {\tt HyRec} code
for the calculation of $X_e$~\cite{AliHaimoud:2010dx}.  The sudden
jump around redshift $z\simeq 1000$ is due to the recombination
effect. At the recombination epoch the ionization fraction $X_e$
decreases dramatically, which leads to a much smaller plasma
mass. Therefore, the probability of the resonant conversion is greatly
enhanced for $z_{\rm res} \lesssim 1000$.  We also give an approximate
formula of the conversion probability as

\begin{align}
	P_{a\to \gamma} \sim 1.7\times 10^{-7}
	\left( \frac{E_0}{1\,\mu{\rm eV}} \right)
	\left( \frac{g_a}{10^{-11}\,{\rm GeV}^{-1}} \right)^2
	\left( \frac{B_0}{1\,{\rm nG}} \right)^2
	\left( \frac{10^{-14}\,{\rm eV}}{m_a} \right)^2
	(1+z_{\rm res})^{7/2}.
	\label{Pres}
\end{align}
In the above expression, we have assumed that the resonance happens in the matter-dominated era
and approximated $r\sim (3H)^{-1}$.  Note that $m_a^2 \propto
(1+z_{\rm res})^3$ and hence Eq.\ (\ref{Pres}) is weakly dependent on
$z_{\rm res}$ as $P_{a\to \gamma} \propto \sqrt{1+z_{\rm res}}$.  Note
also that such low-energy photons may experience bremsstrahlung
absorption above the redshift $z \gtrsim 1700$~\cite{Chluba:2015hma},
and hence the resonant redshift may be constrained as $z_{\rm res}
\lesssim 1700$.

We have made several assumptions for deriving the conversion
probability.  One is that the oscillation length is shorter than the
typical coherent length of the magnetic field.  The oscillation length
is given by
\begin{align}
	\ell_{\rm osc} &\sim \left.\frac{4E}{|\omega_p^2-m_a^2|}\right|_{z=z_{\rm res}}\sim \left.\frac{4\sqrt{Er}}{m_a}\right|_{z=z_{\rm res}}  \nonumber\\
	&\sim 70 \,{\rm kpc}\left( \frac{10^{-14}\,{\rm eV}}{m_a} \right)\left( \frac{E_0}{1\,\mu{\rm eV}} \right)^{1/2}
	(1+z_{\rm res})^{-1/4},
\end{align}
where we have assumed that the resonance happens in the
matter-dominated era and approximated $r\sim (3H)^{-1}$ in the last line.
The coherent length of the magnetic field is assumed to be~\cite{Durrer:2013pga}
\begin{align}
	\ell_B \sim 1\,{\rm Mpc}\, (1+z)^{-1}. % \simeq 1.6\times 10^{29}\,{\rm eV^{-1}} (1+z)^{-1}.
\end{align}
The other condition is that the oscillation length is shorter than the mean free path of the photon, 
\begin{align}
	\ell_{\gamma} = (\sigma_{\rm T} n_e)^{-1} = \frac{3 m_e^2}{8\pi \alpha^2 n_e}
		\simeq 2\times 10^{6}\,{\rm Mpc}\ (1+z)^{-3} X_e^{-1}.
%	\simeq 3.0\times 10^{35}\,{\rm eV}^{-1} (1+z)^{-3} X_e^{-1}.
\end{align}
where $\sigma_{\rm T}$ denotes the cross section of Thomson scattering.
For the parameters of our interest $20\lesssim z_{\rm res} \lesssim 1700$, 
$\ell_{\rm osc}$ is typically smaller than $\ell_B$ and $\ell_\gamma$ and hence we can use (\ref{P_res}) as a conversion probability.

%%%%%%%%%%%%%%%%%%%%%%%%%%%%%%%%%%%%%
\section{Effect of ALP-photon conversion on the CMB}
\label{sec:spec}
%%%%%%%%%%%%%%%%%%%%%%%%%%%%%%%%%%%%%

\begin{figure}[t]
\begin{center}
\includegraphics[width=0.65\textwidth,height=0.5\textwidth]{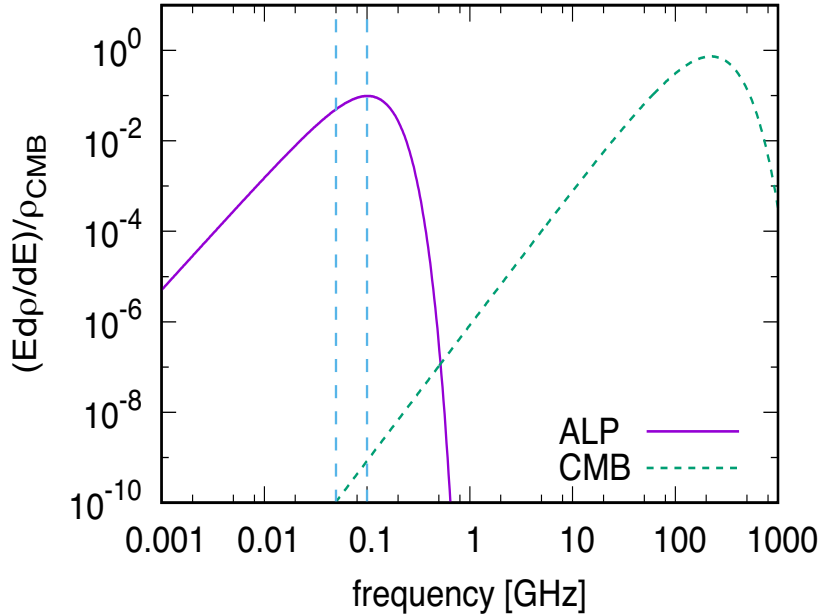}
\caption{An example of the present ALP dark radiation spectrum
  compared with the CMB as a function of the ALP/photon frequency.
  The vertical axis corresponds to the energy spectrum $E d\rho_a/ dE$
  or $E d\rho_\gamma/ dE$ normalized by the total CMB energy density.
  Two vertical lines show the EDGES frequency range.  We have taken
  $m_\phi=40\,\mu$eV, $\tau_\phi = 10^{15}$\,sec and $Y_\phi = 22$ for
  illustration.    \label{fig:spec}}
\end{center}
\end{figure}

Let us consider a process that a moduli-like scalar field $\phi$
decays into the ALP pair, $\phi\to aa$ and assume that the interaction
of ALP is weak enough so that ALPs are regarded as free particles.
The ALP number density spectrum at the redshift $z$ is given as
\begin{align}
	E \frac{dn_a}{dE}(z) 
	%&= E\int \frac{dz'}{(1+z')H(z')}\frac{n_\phi(z')}{\tau_\phi}\frac{a^3(z')}{a^3(z)} \frac{dE'}{dE}2\delta\left(E'-\frac{m_\phi}{2}\right) \nonumber\\
	=  \frac{2n_\phi(z_i) a^3(z_i)}{\tau_\phi H(z_i) a^3(z)},
\end{align}
where $m_\phi$ and $\tau_\phi$ are the mass and lifetime of $\phi$
respectively, $E' = E (1+z') / (1+z)$ and $z_i$ is the redshift at the
production of the ALP with energy $E$ at $z$, which is given by
$1+z_i= m_\phi(1+z)/(2E)$.  The number density of $\phi$ is given by
\begin{align}
	n_\phi(z) = Y_\phi \,s(z) \exp\left( - \frac{t(z)}{\tau_\phi} \right),
\end{align}
where $Y_\phi$ parametrizes the number density of $\phi$.
The total ALP energy density is given by
\begin{align}
	\rho_a(z) = \int dE E \frac{dn_a}{dE}(z).
\end{align}
The present CMB spectrum arising from the conversion of ALP is given
by
\begin{align}
	E_0\frac{dn_\gamma}{dE_0} = \left( E \frac{dn_a}{dE} \right)_{z=z_{\rm res}} 
	P_{a\to \gamma}(z_{\rm res}) \left(\frac{1}{1+z_{\rm res}}\right)^3
	= \left( E \frac{dn_a}{dE} \right)_{z=0} P_{a\to \gamma}(z_{\rm res}).
\end{align}
An example of the present ALP dark radiation spectrum compared with
the CMB (without any absorption and emission by the 21cm line) as a
function of the ALP/photon frequency is shown in Fig.~\ref{fig:spec}.
The vertical axis corresponds to the energy spectrum $E d\rho_a/ dE$
or $E d\rho_\gamma/ dE$ normalized by the total CMB energy density.
Two vertical lines show the EDGES frequency range.  We have taken
$m_\phi=40\,\mu$eV, $\tau_\phi = 10^{15}$\,sec and $Y_\phi = 22$.  We
have $\Delta N_{\rm eff} \simeq 0.26$ (see below) in this parameter
set.

%%%%%%%%%%%%%%%%
%%%%%%%%%%%%%%%%

Now let us estimate the required conversion probability for explaining the EDGES anomaly.
The energy fraction of the CMB in the EDGES frequency range is
\begin{align}
	f_\gamma^{\rm (EDGES)} \equiv  \frac{ \pi^{-2}\int T_0E^2 dE }{ \pi^2 T_0^4/15} \simeq 2.5\times 10^{-10},
\end{align}
where $T_0$ is the present CMB temperature and the integration range is $E = (0.2-0.4)\,\mu$eV.
On the other hand, the energy fraction of the ALP dark radiation in the EDGES frequency range is
\begin{align}
	f_a^{\rm (EDGES)} \equiv  \frac{ \int dE E dn_a/dE}{\rho_a},
\end{align}
where the integration range in the numerator is again $E = (0.2-0.4)\,\mu$eV.
Of course $f_a^{\rm (EDGES)}$ depends on the position of the peak ALP energy.
Numerically we find that it takes a maximum value $f_a^{\rm (EDGES)} \sim 0.4$ when the peak energy is $\sim 0.7\,\mu$eV.
Therefore, in order to increase the photon energy density in the EDGES range by an amount of $\mathcal O(1)$ due to the ALP-photon conversion, we need
\begin{align}
	\rho_a f_a^{\rm (EDGES)} P_{a\to\gamma}(E_0) \sim  \rho_\gamma f_\gamma^{\rm (EDGES)},
\end{align}
with $E_0$ being the EDGES energy range. Thus the required conversion probability is given by
\begin{align}
	P_{a\to\gamma}(E_0) \sim 1.1\times 10^{-9} \,\left(f_a^{\rm (EDGES)} \Delta N_{\rm eff}\right)^{-1},
\end{align}
where we have parameterized the ALP energy density by the extra
effective number of neutrino species $\Delta N_{\rm eff}$, which is
given by
\begin{align}
	\Delta N_{\rm eff} \simeq \frac{\rho_a}{0.23 \rho_{\gamma}},
\end{align}
with $0.23\rho_\gamma$ corresponding to the neutrino energy density of
one species.  The Planck constraint is $\Delta N_{\rm eff} \lesssim
0.33$~\cite{Ade:2015xua}.\footnote{ If ALP dark radiation is
  produced well after the recombination, larger dark radiation energy
  density may be allowed. Also, the upper limit on $\Delta N_{\rm
    eff}$ depends on the cosmological models and some scenarios still
  allow $\Delta N_{\rm eff}\lesssim 0.7$~\cite{Ade:2015xua}.  }
Therefore we need at least $P_{a\to\gamma}(E_0) \gtrsim 8\times
10^{-9}$ at the EDGES frequency range.

There are several constraints on the parameters.  The CAST experiment
is searching for axions from the Sun and the current constraint is
$g_a \lesssim 6.6\times 10^{-11}\,{\rm
  GeV}^{-1}$~\cite{Anastassopoulos:2017ftl}.  Constraint from the
cooling of horizontal branch star is comparable.  The CMB observation
constrains on the magnetic field on the Mpc scale as $B_0 \lesssim$ a
few nG~\cite{Durrer:2013pga,Ade:2015cva}.  The primordial magnetic
field also affects the spin temperature through the extra heating
processes due to some dissipative effects on the
plasma~\cite{Sethi:2004pe,Schleicher:2008hc,Sethi:2009dd,Chluba:2015lpa}.
It may heat up the gas too much for $B_0 \sim 1$\,nG, but the effect
much depends on the shape of power spectrum of the primordial magnetic
field.  The most important constraint in our scenario comes from the
inverse conversion process: CMB photon conversion into the ALP.  Since
the conversion rate is proportional to $E$, it can be important at the
CMB peak frequency range.  Ref.~\cite{Mirizzi:2009nq} derived
constraint on the combination $g_a B_0$ from the distortion of the CMB
due to the conversion of CMB photon into the ALP. The constraint reads
$g_a B_0 \lesssim 10^{-13} - 10^{-11}\,{\rm GeV^{-1}nG}$ for $m_a \sim
10^{-14}\,{\rm eV} - 10^{-9}\,{\rm eV}$ from the measurement by the
COBE FIRAS experiment~\cite{Fixsen:1996nj}.  The ARCADE2 experiment
also measured the CMB flux above the frequency 3\,GHz, although it
does not give a stringent bound compared with the
above~\cite{Fixsen:2009xn}.

\begin{figure}[t]
\begin{center}
\includegraphics[width=0.75\textwidth, height=0.55\textwidth]{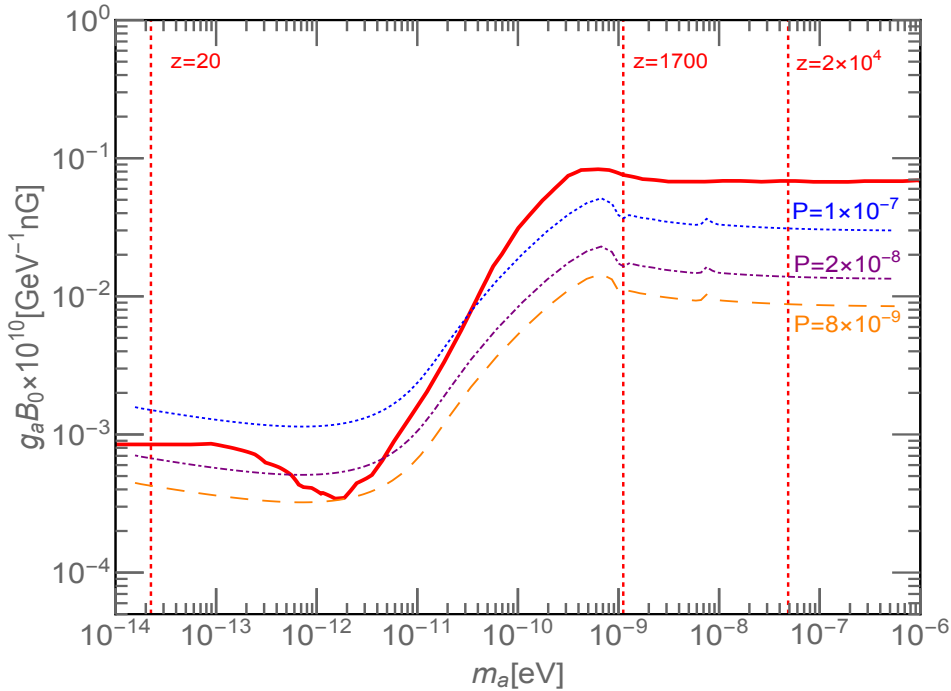}
\caption{Contours for conversion probability $P_{a\to\gamma} = 8\times 10^{-9}, 2\times 10^{-8}$ and $1\times 10^{-7}$, shown in orange long-dashed, purple dot-dashed and blue dotted lines, respectively. The red solid line marks the upper limit on $g_a B_0$ from the CMB spectral distortion~\cite{Mirizzi:2009nq}. 
	Three dashed vertical lines indicate the redshifts at the resonant conversion, $z_{\rm res}=20, 1700,$ and $2\times10^4$, respectively. \label{fig:gaB}}
\end{center}
\end{figure}

In Fig.~\ref{fig:gaB}, on the $(m_a, g_aB_0)$ plane, we show the
contours of constant conversion probability $P_{a\to\gamma}$ for
$E_0\simeq 0.3\,\mu$eV as well as the constraint from distortion of
the CMB spectrum.  The long-dashed orange line gives the conversion
probability $P_{a\to\gamma}\gtrsim 8\times 10^{-9}$, while the purple
dot-dashed and blue dotted ones give $P_{a\to\gamma}=2\times 10^{-8}$
and $1\times 10^{-7}$, respectively. The red solid line is the upper
bound on the product $g_aB_0$ from the CMB spectral distortion given
in Ref.~\cite{Mirizzi:2009nq}.  We also indicate the redshifts at the
resonant conversion $z_{\rm res}=20,1700$ and $2\times 10^4$ with
vertical dashed lines.  To avoid the efficient absorption of the
converted photons, we may need $z_{\rm res} \lesssim
1700$~\cite{Chluba:2015hma} as mentioned earlier.  It is seen that
there are parameter regions in which $P_{a\to\gamma} \gtrsim 10^{-8}$,
which is required for explaining the EDGES anomaly.

%%%%%%%%%%%%
%%%%%%%%%%%%

So far we have not specified the origin of $\phi$.  A general
discussion about the possibility of $\phi$ is beyond the scope of this
paper.  Here, we propose one viable scenario, a simple one based on a
supersymmetric model; we consider a scenario in which the ALP, a
pseudo Nambu-Goldstone boson, is embedded into a complex scalar field
and the radial component of the complex scalar field plays the role of
$\phi$.  (Thus, $\phi$ is also called as ``saxion'' hereafter.)  For
simplicity, we assume that the potential of $\phi$ is well
approximated by a parabolic one, which is the case in a large class of
supersymmetric model.  The decay rate of $\phi$ into the ALP pair may
be given by~\cite{Chun:1995hc}
\begin{align}
  \tau_\phi^{-1}=\Gamma_{\phi\to 2a} = \frac{1}{64\pi}\frac{m_\phi^3}{f^2},
\end{align}
where $f$ is the associated symmetry breaking scale, which is expected
to be related to $g_a$ as $g_a\sim \alpha/16\pi f$. The parent particle
$\phi$ mostly decays at the epoch at $H\sim \Gamma_{\phi\to2a}$.  The
ALP has a energy of $m_\phi/2$ at the production and it is red-shifted
to the EDGES frequency.  We can derive a consistency condition as
\begin{align}
  m_\phi \sim \begin{cases}
    4\times 10^{3}\,{\rm GeV} \left( \frac{f}{10^8\,{\rm GeV}} \right)^2 \left( \frac{1\,\mu{\rm eV}}{E_{\rm peak}} \right)^2 
    & {\rm if}~~\Gamma_{\phi\to 2a} < H_{T=T_{\rm R}}\\
    2\times 10^4\,{\rm GeV} \left( \frac{f}{10^9\,{\rm GeV}} \right)^{4/3} \left( \frac{1\,\mu{\rm eV}}{E_{\rm peak}} \right)
    \left( \frac{T_{\rm R}}{10^3\,{\rm GeV}} \right)
    & {\rm if}~~\Gamma_{\phi\to 2a} > H_{T=T_{\rm R}}
  \end{cases},  \label{mphi}
\end{align}
where $T_{\rm R}$ denotes the reheating temperature and $E_{\rm peak}$
denotes the peak energy of the present ALP spectrum.  Here we simply
assume $m_\phi/2=E_{\rm peak}(1+z_{\rm dec})$ with $z_{\rm dec}$ being
the redshift at the cosmic time comparable to the lifetime of $\phi$,
i.e., $H(z_{\rm dec})=\Gamma_{\phi\to 2a}$.  For the case of low
reheating temperature so that $m_{\phi} > H_{T=T_{\rm R}}$,
for example, the abundance of $\phi$ in the form of coherent
oscillation is given by
\begin{align}
  m_\phi Y_\phi = \frac{1}{8}T_{\rm R}
  \left( \frac{\phi_i}{M_{\rm P}} \right)^2 
  \simeq 1.3\times 10^2\,{\rm GeV}
  \left( \frac{T_{\rm R}}{10^3\,{\rm GeV}} \right) 
  \left( \frac{\phi_i}{M_{\rm P}} \right)^2,
\end{align}
where $\phi_i$ is the initial amplitude of $\phi$ and $M_{\rm P}$ the
reduced Planck scale.  Choosing $\phi_i$ appropriately (close to
$M_{\rm P}$), it is possible to yield ALP with $\Delta N_{\rm eff}
\sim \mathcal O(0.1)$ by the decay of $\phi$.\footnote{The saxion can
  have amplitude much large than $f$ in a supersymmetric
  setup~\cite{Kasuya:1996ns}. There is also a thermal contribution to
the abundance of $\phi$~\cite{Salvio:2013iaa}, but it is subdominant
compared with coherent oscillation if $\phi_i$ is close to $M_{\rm
  P}$.} In order for the momentum distribution of produced ALP not to
be modified by the scattering with thermal photons, it is necessary to
satisfy $\alpha g_a^2 T^3 \lesssim H$, which implies, in the
radiation-dominated universe,
\begin{align}
  T\lesssim 
  10^4\,{\rm GeV} \times
  \left(\frac{g_a}{10^{-10}\,{\rm GeV}^{-1}}\right)^{-2}.
  \label{T_bound}
\end{align}
Based on Eq.\ (\ref{mphi}), we may adopt $f\sim 10^8\,{\rm GeV}$ (and
$g_a\sim 10^{-12}\,{\rm GeV}^{-1}$), $m_\phi\sim 10^3\,{\rm GeV}$, and
$T_{\rm R}\sim 10^3\,{\rm GeV}$, with which the constraint
(\ref{T_bound}) is satisfied while $P\gtrsim 10^{-8}$ is possible
taking $B_0\gtrsim 0.1-1\,{\rm nG}$ to produce enough amount of
radiation in the Rayleigh-Jeans tail.

Before closing this section, let us discuss possibilities to test our
scenario with future experiments.  Interestingly, the future axion
helioscope experiment IAXO~\cite{Irastorza:2011gs} can reach the
sensitivity of $g_a =$(a few)$\times 10^{-12}\,{\rm GeV}^{-1}$.  Thus,
preferred region to explain the EDGES result may be covered.
Future CMB experiments such as PIXIE~\cite{Kogut:2011xw} or
PRISM~\cite{Andre:2013nfa} will also significantly improve the
constraint from CMB spectral distortion, which may confirm or rule out
this scenario.  In the present scenario, $\Delta N_{\rm eff}$ is
required to be larger than $\sim 10^{-2}$; future improvement of the
constraints on $\Delta N_{\rm eff}$~\cite{Abazajian:2016yjj} provides another test of the
present scenario. 

We would also like to compare our scenario with the one proposed in
Ref.~\cite{Pospelov:2018kdh}. Although both scenarios introduce some
hypothetical particle (i.e., ALP or dark photon) that can convert into
photon, the involved physics is different, which would give
distinguishable signals that can be probed by future experiments. In
particular, the transition probability of dark photon to photon is
inversely proportional to the energy~\cite{Pospelov:2018kdh}, while
for ALP it is proportional to the energy (see Eq.~\eqref{P_res}). As a
result, the expected modifications to CMB spectrum would be quite
different.\footnote{For future constraints on dark photon and ALP due to the
  spectral distortion of the CMB, see Refs.~\cite{Kunze:2015noa} and \cite{Tashiro:2013yea}, respectively.}  As
long as future experiments have sensitivity on the relevant energy
range, it should be possible to distinguish those two scenarios by
comparing CMB spectral distortion.

%%%%%%%%%%%%%%%%%%%%%%%%%%%%%%%%%%%%%%%%%%%%
\section{Conclusions}  \label{sec:conc}
%%%%%%%%%%%%%%%%%%%%%%%%%%%%%%%%%%%%%%%%%%%%

Motivated by the recent anomalous enhanced absorption of 21cm radio
signal reported by the EDGES experiment, we have explored a possible
explanation by the ALP-photon resonant conversion. In this scenario,
ALP has a mixing with photon and can convert to photon under
intergalactic magnetic field, which effectively increases the
brightness temperature at the radio band. As long as the resonant
conversion occurs before the redshift $z\gtrsim 20$, the intensity of
the observable 21cm signal can be enhanced relative to the purely
astrophysical effects.  Our scenario does not suffer from the
difficulties faced by other proposals with dark matter scattering with
baryons~\cite{Barkana:2018lgd, Barkana:2018qrx, Berlin:2018sjs,
  Munoz:2018jwq}. Instead, the viable parameter region of ALP-photon
coupling has a strength that can be tested in future axion experiment
such as IAXO.  Future CMB experiments such as PIXIE and PRISM will
significantly improve the bound from CMB spectral distortion, which
may confirm or rule out our scenario.  This scenario may also leave
characteristic signatures on the fluctuation of 21cm lines depending
on the power spectrum of the primordial magnetic field. We leave it as
a future work.

So far we have assumed a pseudo-scalar $a$ that has a coupling to the
photon through $a F_{\mu\nu}\widetilde F^{\mu\nu}$. Most discussion is
parallel for the case of a light scalar field $\sigma$ that has a
coupling through $\sigma F_{\mu\nu} F^{\mu\nu}$; dark radiation
consisting of $\sigma$ may be converted into photons in a similar
fashion under the intergalactic magnetic field.

%%%%%%%%%%%%%%%%%%%%%%%%%%%%%%%%%%%%%%%%%%%%
\section*{Acknowledgments}
%%%%%%%%%%%%%%%%%%%%%%%%%%%%%%%%%%%%%%%%%%%%

We would like to thank T.~Takahashi for pointing out the effect of
magnetic field on the spin temperature.  This work was supported by
the Grant-in-Aid for Scientific Research C (No.\ 18K03608 [TM] and
No.\ 18K03609 [KN]), and Innovative Areas (No.\ 16H06490 [TM and YT],
No.\ 26104009 [KN], No.\ 15H05888 [KN], No.\ 17H06359 [KN]).

%%%%%%%%%%%%%%%%%%%%%%%%%%%%%%%%%%%%%%%%%%%%%%%%%%

%%%%%%%%%%%%%%%%%%%%%%%%%%%%%%%%%%%%%%%%%%%%%%%%%%

\end{document}